\documentclass[review]{elsarticle}

\usepackage{lineno,hyperref}

\newcommand{\sqsn}{$\sqrt{s_{_{NN}}}$}

\newcommand{\bef}{\begin{figure}}
\newcommand{\eef}{\end{figure}}
\newcommand{\bc}{\begin{center}}
\newcommand{\ec}{\end{center}}

\usepackage[usenames,dvipsnames]{color}

\begin{document}

\begin{frontmatter}

\title{Thermal properties of hot and dense medium in interacting hadron 
resonance gas model}

\author[addr1,addr2]{S. Sahoo}
\author[addr3]{D. K. Mishra\fnref{e2}}
\fntext[e2]{dkmishra@barc.gov.in}
\author[addr2]{P. K. Sahu\fnref{e3}}
\fntext[e3]{pradip@iopb.res.in}
\address[addr1]{Siksha `O' Anusandhan Deemed to be University, Bhubaneswar 751030, India}
\address[addr2]{Institute of Physics, HBNI, Sachivalaya Marg, Bhubaneswar, Odisha 751005, India}
\address[addr3]{Nuclear Physics Division, Bhabha Atomic Research Center, Mumbai 400085,India}

\begin{abstract}
 The meson exchange interaction based on relativistic mean-field (RMF) theory has 
been introduced in the hadron resonance gas (HRG) model, called interacting HRG 
(iHRG) model. This model can be used to explain the experimental data both at finite 
temperature ($T$) with finite chemical potential ($\mu_B$) and finite temperature at 
vanishing chemical potential. The nuclear matter equation of state also can be 
explained at zero temperature with finite baryon density (finite chemical potential) 
 due to the presence of attractive and repulsive interactions between the hadrons in 
the iHRG model. Similarly, the lattice equation of state is well described at  
$\mu_B$ = 0 and finite temperature by the iHRG model. In the present study, we have 
calculated the thermodynamical quantities as a function of temperature and chemical 
potential using both HRG and iHRG models. Also, we have presented the isothermal 
compressibility ($k_T$), specific heat ($C_V$), and speed of sound ($c_s^2$) as a 
function of $\mu_B$, $T$, and center of mass energies. The effect of kinematic 
acceptance on these quantities are also presented as a function of $\mu$ and $T$. 
Results from this study on $k_T$ are compared with results from other heavy-ion 
transport models and experimental data up to LHC energies.
\end{abstract}


\end{frontmatter}

\section{Introduction}
\label{intro}
It is expected in nature that with the increase of temperature or baryon 
density to certain high values in the well-defined system, the system undergoes 
a change of its physical properties. This change of physical properties 
is well known as a change of phase transition. 
{Such a phase transition may occur in the strongly interacting 
system such as  nuclear matter and quantum chromodynamics (QCD) 
matter~\cite{Stephanov:2004wx}.}

Since last two decades, it is a matter of interest that a strongly interacting 
matter with increasing temperature and baryon density undergoes a phase transition 
from hadronic to partonic phase, called quark-gluon plasma (QGP) 
state~\cite{aoki2006}. To investigate this finding, there are several well 
established theoretical calculations and experiments on relativistic and 
ultra-relativistic heavy-ion collisions at very high energies or some are 
planned for the near future. There are two extreme limits to have phase transition 
from  hadronic matter to partonic matter. These are at a very small temperature with 
very high baryon densities (in neutron stars)~\cite{sahu01,sahu02} and at small 
baryon density with very high temperature (at Relativistic Heavy-Ion Collider (RHIC) 
and Large Hadron Collider (LHC) energies). Another possibility of such phase 
transition is at moderate temperature and baryons density (Facility for Antiproton 
and Ion Research (FAIR) energies). At small or vanishing chemical potential or very 
low baryon density, the lattice QCD calculations suggest a phase transition from 
hadronic to partonic degrees of freedom is a smooth crossover~\cite{Borsanyi, 
Ding2015}. The crossover phase transition might transform to first-order phase 
transition by passing through a critical end point at finite chemical potential or 
finite baryon density. Besides the lattice QCD~\cite{Senger2011} calculations, there 
are other QCD based models~\cite{vov19,costa09} at finite chemical potential or 
finite baryon density, which are well established to explain the phase transition 
from hadronic to partonic matter and well explain the experimental data from 
heavy-ion collision experiments.

On the experimental front, there are several experiments performed at RHIC and LHC 
energies to study the evidence of the QCD phase transition either directly or 
indirectly from the physical observables. Of course, there are indirect signatures of 
phase transition at the small chemical potential with high temperatures at RHIC and 
LHC energies. But to observe the evidence of this phase transition at higher chemical 
potential and moderate temperature, new experiments have been planned at the 
FAIR~\cite{Ablyazimov:2017guv}
and the Nuclotron-based Ion Collider fAcility (NICA)~\cite{Kekelidze:2016wkp},  
which will start taking data in the next couple of years.

The hadron-resonance gas (HRG) model is very successful to describe physical 
observables from relativistic heavy-ion collisions at AGS to RHIC and LHC 
energies~\cite{Vov2015,Sam2018}. The basic concept of the model is that the medium 
created by heavy-ion collisions start equilibrating with time evolution and continue 
to interact until it reaches to the freeze-out limit. In this model, the hadrons are 
treated as non-interacting particles. The HRG model successfully describes the 
particle ratios at finite chemical potential in both canonical and grand canonical 
approaches. The HRG model reproduce the results of lattice QCD  at zero 
baryon-chemical potential with finite temperature ($T\sim$ 170 MeV).  Also, the HRG 
model at finite temperature and chemical potential is very successful in describing 
the experimental data at RHIC and LHC energies. However, this model fails to 
describe the nuclear matter properties at finite chemical potential and vanishing or small 
temperature, such as the nuclear hadronic equation of state at saturation density but 
works very well at vanishing chemical potential at finite temperature, such as 
lattice QCD. The reason for failing  the HRG model at around nuclear matter
saturation regions, because it does not have repulsive and attractive interaction
between the hadrons, which are very important when there is interaction in the
medium and helps to get saturation properties at nuclear matter. The HRG model
works very well at small $\mu_B$ and higher $T$, because the matter leads to an
ideal gas of satisfying the thermodynamics description of the particles up
to $T\sim 170$ MeV.

To overcome the shortcomings of the HRG model at finite chemical potential or finite 
baryon density, we have included the repulsive and attractive interaction 
between baryons in the HRG model through the relativistic mean-field (RMF)
theory~\cite {datta93,sahu93}. Recently, an attempt has been made to include
interactions between baryons based on the RMF theory in the HRG model~\cite{cass18} 
and repulsive  baryon–baryon interactions in a hadron gas model~\cite{Huovinen:2017ogf}.
The RMF models are very good to explain the properties of infinite nuclear matter,
such as the gross structure of neutron stars~\cite{sahu00} and properties of finite
nuclei~\cite{sahu10}, as the scalar interaction is for attractive and vector interaction
is for repulsive potentials, which are included through the meson exchange interactions
with proper parameters. However, these 
models fail to explain the matter at low density or vanishing chemical potential and at 
finite high temperatures, such as QCD matter. Hence, in the RMF models the QCD 
matter properties such as specific heat, susceptibility, thermal compressibility and
equation of state at finite temperature are not well described.

In view of the above reasons, we adopt a combination of both RMF and HRG models, 
which we call as iHRG model. Therefore, it may explain both vanishing chemical 
potential to finite chemical potential with zero to finite temperature i.e. hadronic 
nuclear matter to QCD matter. In the present study, we consider RMF chiral sigma 
model~\cite{sahu93,sahu10,Papazoglou:1998vr,sahu00a,sahu04}, which can describe well the ground state 
properties of infinite nuclear matter, neutron star gross properties and properties 
of finite nuclei. Along with this model, we combine the HRG model to explain the 
lattice QCD and heavy-ion collision data at RHIC and LHC at small/vanishing chemical 
potential and at finite temperature near hadronic to a partonic phase transition.

The paper is organized as follows. In the following sections II and III, we describe 
the ideal HRG model and interacting HRG (iHRG) model, with subsections of RMF theory 
with zero and finite temperature, respectively. The results and discussions are given 
in section IV and finally, we summarize our findings in section V.

\section{Hadron resonance gas model}
The HRG model is based on the work of Dashen, Ma, and Bernstein~\cite{Dashen:1969ep},
which shows that one can describe the strongly interacting system by a gas of free
hadrons and inclusion of resonances in a thermal medium. Further these resonances can be related to phase shift~\cite{Lo:2017sde,Cleymans:2020fsc}. In this model,
the stable hadrons and the resonances are considered as a non-interacting point
particles. This simple statistical model is successful in describing the particle
abundances in nucleon-nucleus~\cite{Sharma:2018owb}, and heavy-ion collisions for
different collision energies typical of Alternating Gradient Synchrotron (AGS) up
to those of the Large Hadron Collider 
(LHC)~\cite{BraunMunzinger:2003zd,Cleymans:2005xv,Andronic:2011yq}. The HRG model is 
also successfully applied to smaller systems such as: $p+p$~\cite{Vovchenko:2015idt}
and $e^+ + e^-$ collisions~\cite{Becattini:2010sk}. 
It has been observed that, ideal HRG model is quite successful in reproducing the Lattice QCD (LQCD) calculations of the 
thermodynamic parameters of QCD matter at $\mu \approx$ 0  and below temperature 
$T\approx$ 170 MeV. However, the disagreement between LQCD and HRG model calculations 
are observed at higher temperatures as well as even at low temperature for 
fluctuations~\cite{Huovinen:2017ogf,Lo:2017lym}.

The heavy-ion experiments cover only a limited phase space, so one can access only a 
part of the fireball produced in the collision, which resembles the grand canonical 
ensemble (GCE)~\cite{Koch:2008ia,Karsch:2010ck}. There is no requirement of 
conservation of energy, momentum, and charge measured in the limited phase-space in 
the GCE model, which is similar to the experimental situation in heavy-ion 
collisions. The thermodynamic potential for the hadronic system is given by the sum 
of the potential of all stable hadrons and all known resonances. Assuming a thermal 
system produced in the heavy-ion collisions, the logarithm of the partition function 
of a hadron resonance gas in the grand canonical ensemble can be written as 
$\mathrm{ln} Z$ = $\sum_i \mathrm{ln} Z_i$, where the sum is over all the stable 
hadrons and resonances. In the ambit of GCE framework, the logarithm of the 
partition function of $i$-th particle is given as 
\begin{equation}
\label{eq:lnz}
\mathrm{ln} Z_i(T, V, \mu_i) = \pm \frac{Vg_i}{(2\pi)^3}\int d^3k
~\mathrm{ln}\big [1\pm \mathrm{exp}\{(\mu_i-E)/T\}\big ],
\end{equation}
where $V$ is the volume of the thermal system, $g_i$ being the degeneracy factor for 
$i$-th particle, $T$ is the temperature and $\mu_i$ is the chemical potential. The 
single-particle energy $E = \sqrt{k^2 + m^2}$, where $m$ is the mass of the particle 
and $k$ being the momentum. The $\pm$ sign corresponds to the fermion and boson 
respectively. All the hadrons having mass up to 2.5 GeV as listed in the PDG are 
considered in the present calculations. Using partition function one can calculate 
various thermodynamical quantities of the thermal system formed in heavy-ion 
collisions. The pressure ($P$), energy density ($\varepsilon$), number density ($n$) 
of the thermal system is defined as:
\begin{equation}
P = \sum_i\pm \frac{g_iT}{2\pi^2}\int_0^\infty k^2 dk~
\ln[1\pm\exp\{(\mu_i-E_i)/T\}],
\end{equation}
\begin{equation}
 \varepsilon = \sum_i\frac{g_i}{2\pi^2}\int_0^\infty 
\frac{k^2dk}{\exp[(E_i-\mu_i)/T]\pm 1}E_i, 
\end{equation}
\begin{equation}
n = \sum_i\frac{g_i}{2\pi^2}\int_0^\infty\frac{k^2dk}{\exp[(E_i-\mu_i)/T]\pm 1}. 
\end{equation}

Using the above equations, one can calculate the other thermodynamical quantities 
such as entropy density ($s$), square of speed of sound ($c_s^2$), specific 
heat ($C_V$), and isothermal compressibility ($k_T$) as follows. The pressure 
and energy density produce at the early stages of the collisions which leads 
to an increase in entropy 
density. Entropy being directly related to the experimentally measured 
charged-particle multiplicity in a given pseudorapidity range is an important 
observable to study the QCD phase transition. In thermodynamics, it is defined 
as
\begin{equation}
 s = \frac{\varepsilon + P -\sum_i \mu_i n_i}{T}.
\end{equation}
In hydrodynamics, the speed of sound plays an important role in understanding the 
equation-of-state and hence, the associated phase transition. It is related to the 
small perturbations produced in the medium formed in heavy-ion collisions. It 
explains the rate of change of pressure gradients due to the change in the energy 
density of the created dense medium. It is defined as
\begin{equation}
 c_s^2 = \frac{d P}{d \varepsilon}.
\end{equation}
The heat capacity or the specific heat ($C_V$) is a thermodynamic quantity 
characterizing the equation of state of the system. It is also related to the change 
in entropy of the thermal system with a change in temperature ($ds = \int\frac{C_V}{T}dT$).
In the case of a system undergoing a phase transition, $C_V$ is 
expected to diverge at the critical point. Close to the critical point, the specific 
heat is expressed in terms of power-law as $C_V \propto |T-T_c|^{-\alpha}$,
where $T_c$ is the critical temperature and $\alpha$ being the critical exponent. The 
specific heat at constant volume is defined as,
\begin{equation}
 C_V = \frac{d\varepsilon}{dT}.
\end{equation} 
Isothermal compressibility ($k_T$) the measure of the relative change in volume 
with respect to change in pressure at a constant temperature. Similar to $C_V$, the 
$k_T$ is also diverge at the critical point and close to the critical point, it can be 
expressed in terms of power-law. Hence, the determination of $C_V$ and $k_T$ may 
help in understanding the critical point and the phase transition and their nature. 
The isothermal compressibility is defined as 
\begin{equation}
 k_T = -\frac{1}{V}\frac{dV}{dP} = \frac{1}{V}\frac{\sum_i 
\frac{dn_i}{d\mu_i}}{\sum_i (\frac{dP}{d\mu_i})^2}.
\label{eq8}
\end{equation}
The above  pressure ($P$) and energy density ($\varepsilon$) may be represented 
as 
($P_{HRG}$) and ($\varepsilon_{HRG}$), respectively for ideal (non-interacting) 
hadron resonance gas model.

\section{ Interacting Model and Interacting HRG model}
The nuclear equation of state (EOS) is very important in nuclear physics 
and astrophysics~\cite{ran93,fai96,zha96,Steinheimer:2010ib}, specially the
liquid-gas phase transition at lower density and finite temperature calculation of 
nuclear many-body system~\cite{subal00,das92}. To study the properties
of quark gluon plasma (QGP) at extreme densities and temperatures~\cite{sahu02a}
and the study of  stellar evolution, the global properties of neutron 
star and supernova explosion~\cite{sahu00,shen98,brown82,Motornenko:2019arp}, 
the nuclear EOS is also needed.

Theoretically, many-body approaches such as Hartee-Fock, Thomas-Fermi and
mean-field theory type procedures~\cite{bonche81,serot86} have been adopted
to derive the EOS. One of the very successful theoretical calculation 
of finite nuclei and infinite nuclear 
matter~\cite{sahu00,sahu10,Papazoglou:1998vr,sahu02a}
is the relativistic mean-field (RMF) theory. To  describe the required values
of saturation properties of nuclear matter, incompressibility, binding energy,
and effective nucleon mass, the original Walecka RMF model~\cite{wal74} has been 
extensively modified. The non-linear term in RMF~\cite{manka} has been introduced
to describe the finite nuclei properties and the nuclear matter 
properties at normal nuclear density, however, it differs from the relativistic 
Dirac-Brueckner-Hartee-Fock (DBHF) EOS~\cite{haar87}. Therefore, there are 
several models , which attempt to reproduce the reasonable EOS, compatible with DBHF, 
by including vector mesons self couplings. 

One such model in the high density nuclear matter~\cite{sahu93} is the chiral
sigma (CS) model. Like RMF model, the CS model is very successful. The meson fields
are calculated similar way based on the mean-field approximation. The non-linear
terms here behave like the three-body forces and are main ingredients to reproduce
the properties of nuclear matter saturation. We had adopted a SU(2) CS model in
the beginning, where the mass of the iso-scalar vector field is generated 
dynamically~\cite{sahu93}, however, high value of incompressibility was the
main shortcoming. To get desired values of compressibility, effective mass,
binding energy, and saturation density, the higher-order terms of the scalar meson 
field~\cite{sahu10,sahu00a,sahu04}, were introduced. 
Using non-linear SU(2) CS model, we evaluate the nuclear EOS at zero and finite 
temperature. In this calculation, we choose the stable parameter STO-5 
from Ref.~\cite{sahu10} and evaluate the thermodynamic quantities, which are  the 
applicable to various heavy-ion collision experiments.

\subsection{The Formalism of Chiral Sigma Model}
The {\it SU}(2) chiral sigma Lagrangian can be written as \cite{sahu93}

\begin{eqnarray}
\label{lag}
{\cal L} &=&  \frac{1}{2}\big(
         \partial_{\mu} {\bf \pi} \cdot \partial^{\mu} {\bf \pi}
        +\partial_{\mu} \sigma \partial^{\mu} \sigma
        \big)
- \frac{1}{4} F_{\mu\nu} F_{\mu\nu}
\nonumber \\
&&- \frac{\lambda}{4}\big(x^2 - x^2_o\big)^2
- \frac{\lambda b}{6m^2}\big(x^2 - x^2_o\big)^3
- \frac{\lambda c}{8m^4}\big(x^2 - x^2_o\big)^4
\nonumber \\
&&- \sum_i g_{i \sigma} \bar{\psi_i} \big(
                \sigma + i\gamma_5 {\bf \tau}\cdot{\bf \pi}
        \big) \psi_i
+ \bar\psi_i \big(
i\gamma_{\mu}\partial^{\mu} - g_{i \omega}\gamma_{\mu}
\omega^{\mu}\big) \psi_i
\nonumber \\
&&+ \frac{1}{2}{g_{N \omega}}^{2}
x^2 \omega_{\mu}
\omega^{\mu} + \frac{1}{24}\xi {g_{N \omega}}^4(\omega_{\mu}\omega^{\mu})^2
\end{eqnarray}
Here $F_{\mu\nu}\equiv\partial_{\mu}\omega_{\nu}-\partial_{\nu}
\omega_{\mu}$ and  $x^2= {\bf \pi}^2+\sigma^{2}$.
In the Lagrangian $\psi_i$ is for nucleons, $\Delta$, $\Lambda$,
$\Sigma^-$ and $\Xi$ hyperons (denoted by subscript $i$), ${\bf \pi}$ is
the pseudoscalar-isovector pion field and $\sigma$ is the scalar field.
We consider in natural units with $ \hbar = c = k_{B} = 1$.

The main point in this Lagrangian is that it has a dynamically generated 
iso-scalar vector field, $\omega_{\mu}$, with conserved baryonic 
current $j_{\mu}=\bar{\psi}\gamma_{\mu}\psi$. The constant parameters $b$ and $c$ 
are included in the higher-order self-interaction of the scalar field in the 
potential to describe the  nuclear saturation density. The modified non-linear CS 
model is defined by STO-5 in our following discussions. In the fourth-order term in 
the omega fields, the quantity $\xi$ is a constant parameter. For simplicity, we set 
$\xi$ to zero. A mass is generated when the interactions of the scalar and 
the pseudoscalar mesons with the vector meson take place through the 
spontaneous breaking of the chiral symmetry.

The masses of the nucleon, the scalar meson and the vector meson are respectively 
given by
\begin{eqnarray}
m = g_{\sigma} x_o,~~ m_{\sigma} = \sqrt{2\lambda} x_o,~~
m_{\omega} = g_{\omega} x_o\ ,
\end{eqnarray}
where $x_o$ is the vacuum expectation value of the $\sigma$ field, 
$\lambda~=~({m_{\sigma}}^{2}-{m_{\pi}}^{2})/(2 {f_{\pi}}^{2})$, with $m_{\pi}$ 
is the pion mass and $f_{\pi}$ the pion decay coupling constant, and $g_{ 
N \omega}=g_{\omega}$ and $g_{N \sigma}=g_{\sigma}$ are the coupling constants
for the vector and scalar fields, respectively. In the mean field approximation,
we assume the pion field to be zero because of it's pseudoscalar in nature.
 However, there is an interaction between nucleon and pion.
 The inclusion of pions in the thermodynamic trace is due to
 $<\delta\pi\delta\pi>$ propagator, with an interacting pion mass
 determined from the RMF model.

By adopting mean-field approximation, the equation of motion of fields is obtained. 
This approach has been used extensively to evaluate the equation of state in any 
theoretical models for high density and finite temperature of the nuclear matter.

The equation of motion for the scalar field is given below based on the ansatz of mean-field 

\begin{eqnarray}
 (1-R^2) -\frac{b}{m^2 c_{\omega}}(1-R^2)^2
+\frac{c}{m^4c_{\omega}^2}(1-R^2)^3 \nonumber \\
+\frac{2 c_{\sigma}c_{\omega}n_B^2}{m^2R^4}
-  \sum_i  { \frac{c_{\sigma}\gamma}{\pi^2}\int^{k_F}_o\frac{k^2dk}
        {\sqrt{k^2+m_i^{\star 2}}}=0\ ,}
\label{effmass}
\end{eqnarray}

where $m^{\star} \equiv R m$ is the effective mass of the nucleon.  The terms  $c_\sigma 
\equiv  g_{\sigma}^2/m_{\sigma}^2 $  and $c_{\omega} \equiv g_{\omega}^2/m_{\omega}^2$
are scalar and vector coupling constants, respectively.

The equation of motion for the iso-scalar vector field is
\begin{equation}
\omega_0=\frac{ g_{\omega} x^2 \gamma}{(2\pi)^3}\int^{k_F}_o d^3k \ .
\end{equation}

The quantity $k_F$ is the Fermi momentum and the degeneracy factor $\gamma$ is the nucleon spin.

\subsection{Hadronic matter at finite temperature}

The equation of state for finite temperature is as follows:

\begin{eqnarray}
\label{ept}
\varepsilon_{RMF}(T)
&=&
          T1
        - T2     
        + T3  
         +T4
\nonumber \\      
 &&    + \frac{\gamma}{2\pi^2} \sum_i 
                \int _o^{\infty} k^2dk\sqrt{{k}^2 + m_i^{\star 2}} 
        (d_i(T)+\bar d_i(T))\ ,
\nonumber\\
P_{RMF}(T) &=&
       -T1+T2-T3+T4
          \nonumber\\
          &&
        + \frac{\gamma }{6\pi^2} \sum_i 
                \int _o^{\infty} \frac{k^4dk}{\sqrt{{k}^2 + m_i^{\star 2}}} 
         (d_i(T)+\bar d_i(T))\ .
\end{eqnarray}
Where, 
\begin{eqnarray}
  T1&=& \frac{m^2(1-R^2)^2}{8c_{\sigma}},
  \nonumber \\
  T2&=&  \frac{b}{12c_{\omega}c_{\sigma}}(1-R^2)^3,
\nonumber \\     
     T3&= & \frac{c}{16m^2c_{\omega}^2c_{\sigma}}(1-R^2)^4,
     \nonumber \\
      T4&=&  \frac{c_{\omega} n_B^2}{2R^2}.
          \end{eqnarray}
          
The baryon and scalar densities at finite temperature are respectively
defined as
\begin{eqnarray}
n_B(T)= \frac{\gamma}{(2\pi)^3}\sum_i \int^{\infty}_o d^3k (d_i(T)-\bar d_i(T)) ,
\nonumber \\
n_S(T)= \frac{\gamma}{(2\pi^)3}\sum_i \int^{\infty}_o\frac{m_i^*  d^3k}
        {\sqrt{k^2+m_i^{\star 2}}} (d_i(T) + \bar d_i(T)) .
\end{eqnarray}

The Eq. \ref{effmass} is modified as 
\begin{eqnarray}
(1-R^2) -\frac{b}{m^2 c_{\omega}}(1-R^2)^2\nonumber\\
+\frac{c}{m^4c_{\omega}^2}(1-R^2)^3 
+\frac{2 c_{\sigma}c_{\omega}n_B^2}{m^2R^4}\nonumber\\
-\sum_i \frac{c_{\sigma}\gamma}{\pi^2}\int^{\infty}_o\frac{k^2dk}
        {\sqrt{k^2+m_i^{\star 2}}}(d_i(T)+\bar d_i(T))=0\ .
\label{effmass1}
\end{eqnarray}

The distributions functions for nucleon $d(T)$ and anti-nucleon $\bar d(T)$, are 
respectively, given as

\begin{eqnarray}
 d_i(T)=\sum_i  \frac{1}{\exp{[(E_i^{\star}+\mu_i^{\star})/T]}+1}  \nonumber \\
\&  \nonumber \\
 \bar d_i(T)=\sum_i \frac{1}{\exp{[(E_i^{\star}-\mu_i^{\star})/T]}+1}.
\end{eqnarray}

where $E_i^{\star} = \sqrt{k^2+m_i^{\star ^2}}$, $T$ is temperature.

Since the Lagrangian above includes nucleons, hyperons, $\sigma$, and  $\omega$ 
mesons, the meson fields interact with baryons through linear coupling. The 
coupling constants are different for non-strange and strange baryons. The 
$\omega$  mass is chosen to be physical masses. The equation of state is 
obtained through the mean field ansatz. In this case, one can define effective 
masses ($m_i^*$) and chemical potentials ($\mu_i^*$) for the baryons as

\begin{eqnarray}
m_i^* & = & - g_{i \sigma} \sigma_0 \nonumber \\
\mu_i^* & = & \mu_i - g_{ i \omega}  \omega_0,
\end{eqnarray}
where $ \omega_o$, and $\sigma_o$ are the non-zero vacuum expectation values of 
the meson fields.

\subsection{Hadronic matter at zero temperature}
The equation of state is calculated from the diagonal components of the 
conserved 
total stress tensor corresponding to the Lagrangian together with the mean-field 
equation of motion for the fermion field and a mean-field approximation for the 
meson fields. The total energy density, $\varepsilon$, and pressure, $P$ of the 
many-nucleon system are defined as the following:

\begin{eqnarray}
\label{ep0}
\varepsilon_{RMF}(0)
&=&
          T1-T2+T3+T4
\nonumber \\
&&
             + \frac{\gamma}{2\pi^2}\sum_i 
                \int _o^{k_F} k^2dk\sqrt{{k}^2 + m_i^{\star 2}}\ ,
\nonumber\\
P_{RMF}(0) &=&
        -T1+T2-T3+T4
\nonumber \\
&&
         + \frac{\gamma }{6\pi^2}
                \int _o^{k_F} \frac{k^4dk}{\sqrt{{k}^2 + m_i^{\star 2}}}\ .
\end{eqnarray}

The energy per nucleon is $E/A=\varepsilon/n_B$, where $\gamma=4$ for symmetric 
nuclear matter.

The baryon  $n_B$  and scalar densities $n_S$ are respectively defined as

\begin{eqnarray}
 n_B= \frac{\gamma}{(2\pi)^3}\sum_i \int^{k_F}_o d^3k,
\nonumber\\
n_S= \frac{\gamma}{(2\pi)^3}\sum_i \int^{k_F}_o\frac{m^*}
        {\sqrt{k^2+m_i^{\star 2}}} d^3k,
\end{eqnarray}

these are used in Eq.(\ref{effmass}).

\subsection{Interacting HRG model}
The thermodynamic quantities and equation of states are calculated using both the 
ideal HRG model as well as interacting HRG (iHRG) model. The energy density and 
pressure are defined in iHRG model as

\begin{eqnarray}
\varepsilon_{iHRG}=\varepsilon_{RMF}+\varepsilon_{HRG}-\varepsilon_{N}-\varepsilon_{\Delta}
\nonumber\\
P_{iHRG}=P_{RMF}+P_{HRG}-P_{N}-P_{\Delta}
\end{eqnarray}
Where $\varepsilon_{N}$, $\varepsilon_{\Delta}$, $P_{N}$ and $P_{\Delta}$ represent 
the energy density and pressure of 
the non-interacting nucleons and deltas in the HRG model. The interaction for all baryons such as 
nucleons ($N$) and Deltas ($\Delta$) are taken as same as nucleons. However, for 
hyperonic baryons, the interaction is taken as described below.

The nonlinear RMF hadronic model has a couple of parameters, which are 
determined by 
the properties of nuclear matter. The nucleon couplings to scalar 
($g_\sigma/m_\sigma$), and vector mesons ($g_\omega/m_\omega$) and the two 
coefficients $b$ and $c$ in Eq. \ref{effmass}, are obtained by fitting saturation 
values of nuclear matter binding energy per nucleon ($\sim-14.5$ MeV) and 
baryon density ($\sim0.14$ fm$^{-3}$), and Landau mass (0.85 $m_N$). We have 
taken the stable set of parameters STO-5, given in Ref.~\cite{sahu10}.
\begin{figure}[!h]
\bc
\includegraphics[width=0.8\textwidth]{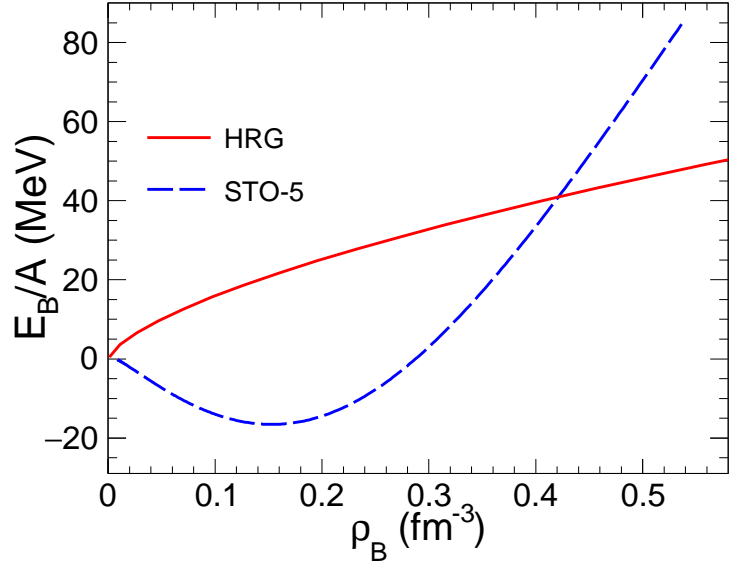}
\caption{The binding energy per nucleon ($E_B/A$) as a function of nucleon 
density ($\rho_B$) at $T=$ 0. The solid red line is the result from the 
ideal HRG model and the dashed blue line is for interacting HRG model.  } 
\label{fig:fig1}
\ec
\end{figure}
The coupling constant parameters of the hyperon (ratio of hyperon-meson and 
nucleon-meson) are uncertain because hyperons do not present in the nuclear matter, 
hence cannot be determined from the nuclear matter properties. Therefore, we choose 
the value of hyperon couplings for scalar and vector mesons as 2/3 (similar to quark 
counting value for $\Lambda$, $\Sigma$ and $\Xi$). The detail is referred to 
Refs.~\cite{sahu01,sahu02} and references therein.

\section{Results and discussion}
\label{sec:results}

\begin{figure}[h]
\bc
\includegraphics[width=0.7\textwidth]{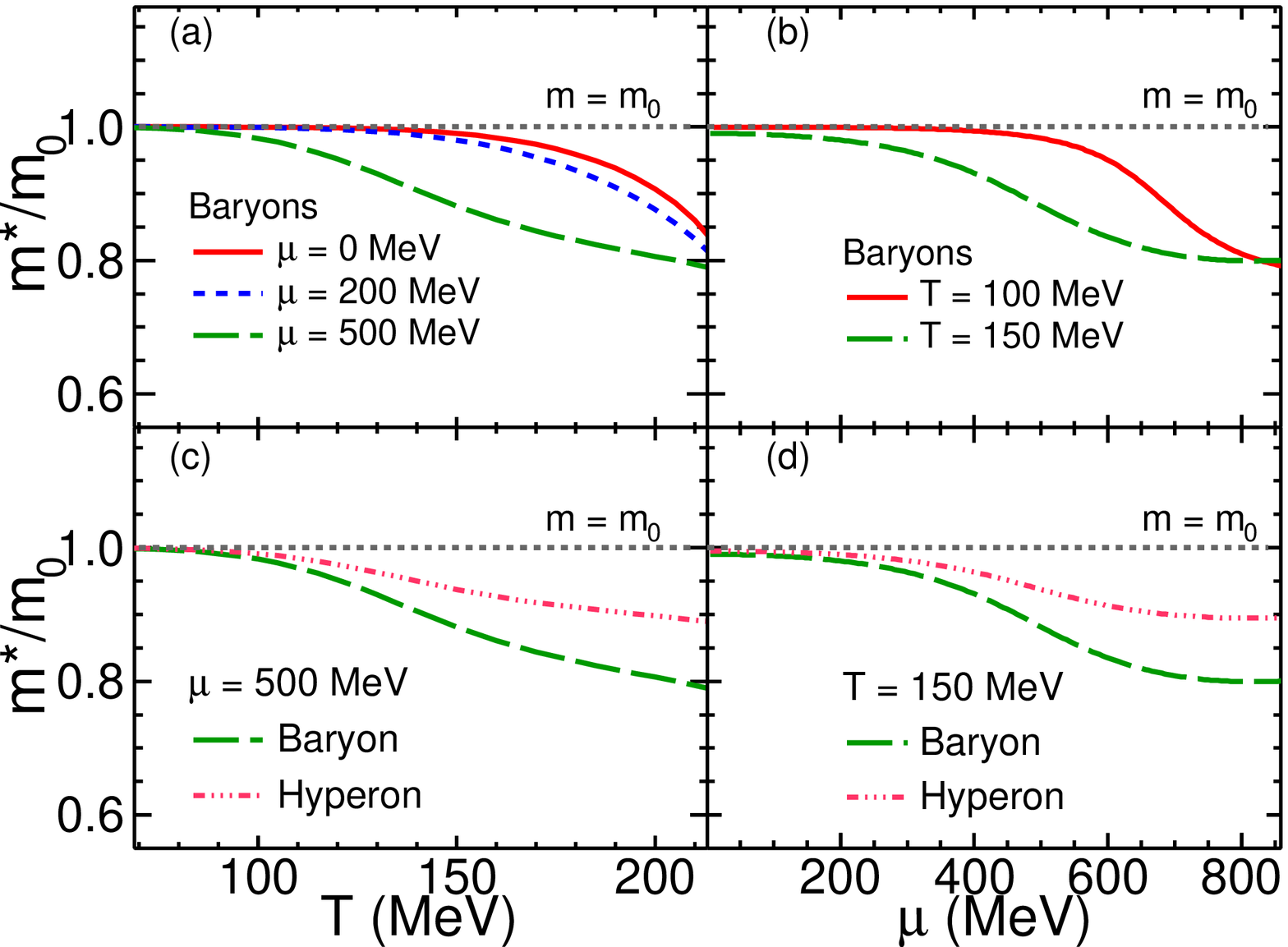}
\caption{Ratios of the effective mass ($m^*$) of the baryons (hyperons) 
relative to their rest mass ($m_0$) as a function of $T$ for difference $\mu$ 
values (panel (a)) and function of $\mu$ for different $T$ values (panel (b)). 
The ratios of effective masses of hyperons as a function of $T$ (panel (c)) and 
function of $\mu$ (panel (d)). } 
\label{fig:fig2}
\ec
\end{figure}

\begin{figure}[!h]
\bc
\includegraphics[width=0.8\textwidth]{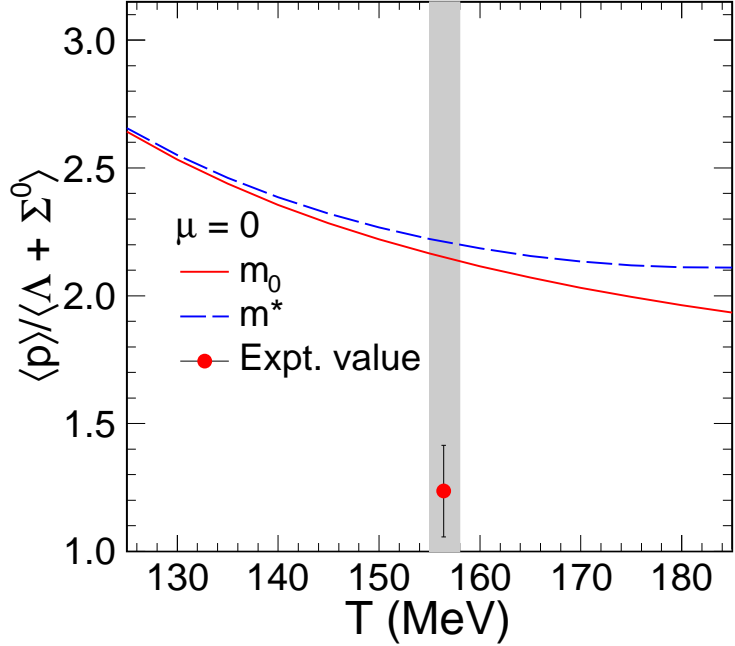}
\caption{The ratio of proton to that of $\Lambda + \Sigma^0$ baryons as a function of $T$. The solid red line is the result from the ideal HRG model and the dashed blue line is for interacting HRG model. The solid symbol depicts the experimental value measured by the ALICE collaboration for the most central Pb-Pb collisions (at $\sqrt{s_{_{NN}}} =$ 2.76 TeV)~\cite{ALICE:2013mez, ALICE:2013cdo, ALICE:2013xmt}.} 
\label{fig:fig2_1}
\ec
\end{figure}

\begin{figure*}[!t]
\bc
\includegraphics[width=1.0\textwidth]{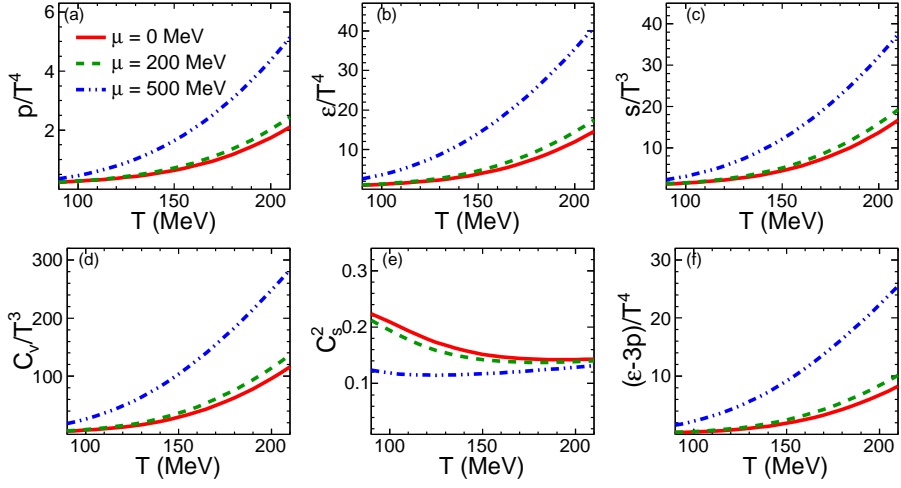}
\caption{The variation of different thermodynamic quantities ($p/T^4$, 
$\varepsilon/T^4$, $s/T^3$, $C_V/T^3$, $c_s^2$, and $(\varepsilon-3p)/T^4$) as 
a function of $T$ for $\mu = $ 0 (soild red line), 200 MeV (dashed green line), 
and 500 MeV (dotted-dashed blue line) calculated using ideal HRG model. } 
\label{fig:fig3}
\ec
\end{figure*}
\begin{figure*}[!ht]
\bc
\includegraphics[width=1.0\textwidth]{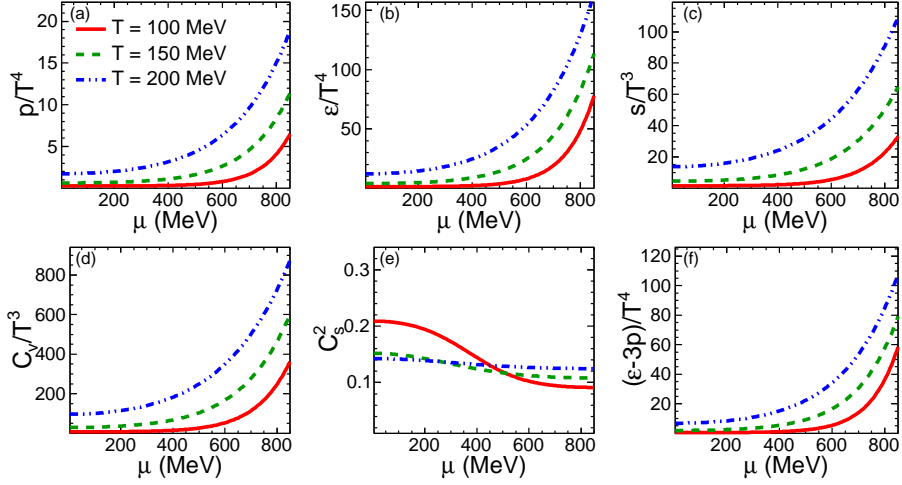}
\caption{The variation of different thermodynamic quantities as shown in 
Fig.~\ref{fig:fig3} as a function of $\mu$ for different values of $T =$ 100 
MeV (solid red line), 150 MeV (dashed green line) and 200 MeV (dotted-dashed 
blue line) using ideal HRG model.} 
\label{fig:fig4}
\ec
\end{figure*}
\begin{figure*}[!ht]
\bc
\includegraphics[width=1.0\textwidth]{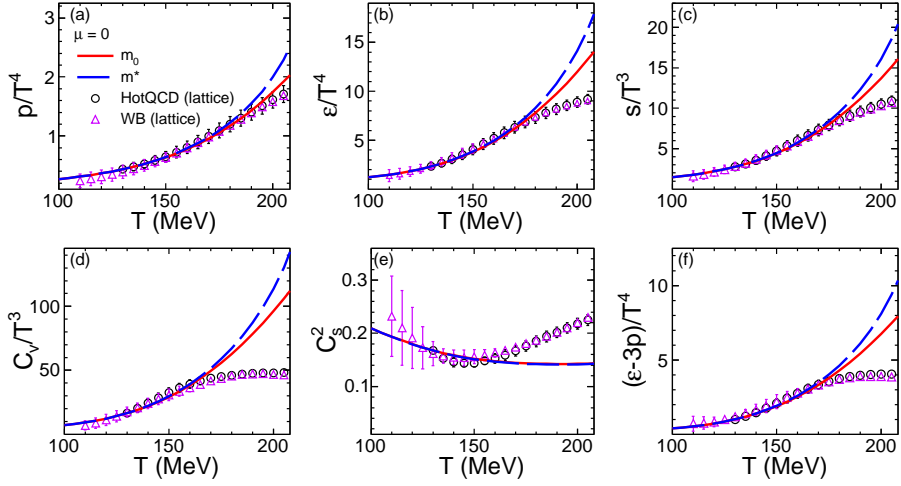}
\caption{ The equation of states of interacting HRG (iHRG) as a function of 
temperature $T$. The solid red line for ideal HRG model (with $m=m_0$) and 
blue dashed line for iHRG (with $m = m^*$) model. The open circle and magenta 
triangle symbols are for HotQCD~\cite{Bazavov:2014pvz} and 
WB~\cite{Borsanyi:2013bia} lattice data respectively. } 
\label{fig:fig5}
\ec
\end{figure*}

\begin{figure*}[!ht]
\bc
\includegraphics[width=1.0\textwidth]{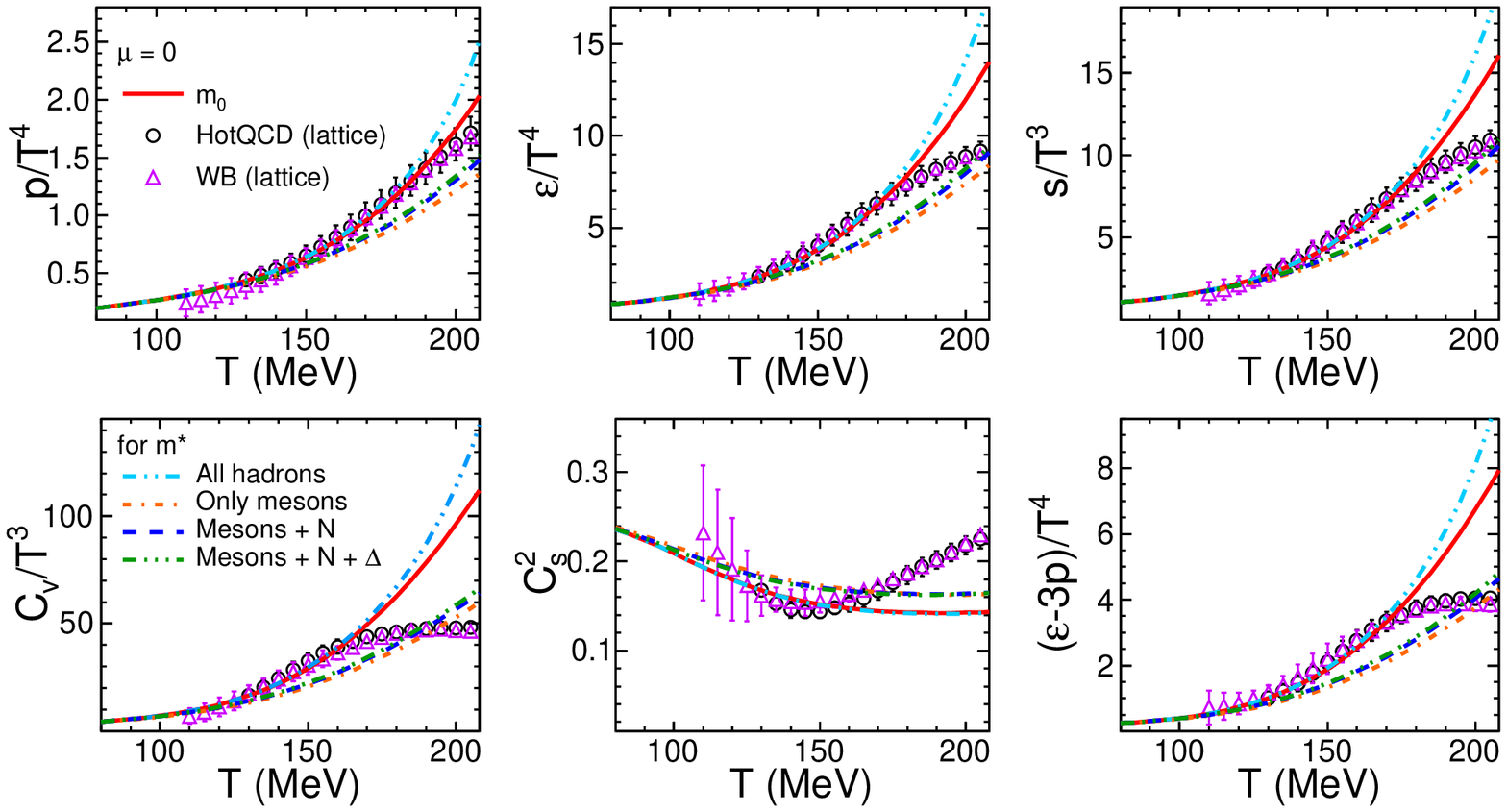}
\caption{ The equation of states of interacting HRG (iHRG) as a function of 
temperature $T$. There are five choices of equation of states: for all 
particles with $m = m_0$ (solid red line), for all hadrons with $m=m^*$ 
(dashed double dots cyan line), for only mesons (dashed dotted orange line), 
mesons with only nucleons (blue dashed line), mesons along with nucleons and 
deltas resonances(green dashed triple dot line). The open circle  and magenta 
triangle symbols are for HotQCD~\cite{Bazavov:2014pvz} and 
WB~\cite{Borsanyi:2013bia} lattice data respectively.} 
\label{fig:fig6}
\ec
\end{figure*}
Figure~\ref{fig:fig1} shows the binding energy as a function of baryon density at 
zero temperature for both HRG and iHRG models. The parameters are considered in such 
a way that the iHRG model can describe the required saturation density and binding 
energy at the ground state of nuclear matter. In the iHRG model, interactions 
between baryons are based on the relativistic mean-field theory. Further, the 
repulsive and attractive interactions are introduced through meson-exchange reaction. 
We have used the same particles in the iHRG model along with all resonance 
particles as used in the HRG model. The solid red line represents the binding 
energy per 
nucleon for non-interacting nucleons with all resonances in the HRG model. The dashed 
blue line shows the equation of state of iHRG used in the HRG model. Since there is 
no interaction among the nucleons in the HRG model, so the binding energy at 
saturation density does not reproduce the ground state value of normal nuclear 
matter. 
Figure.~\ref{fig:fig2} shows the ratios of effective masses to the bare masses as a 
function of temperature at different $\mu$ (Fig.~\ref{fig:fig2}a) and as a function 
of baryon chemical potential at different $T$ (Fig.~\ref{fig:fig2}b). The dotted 
lines shown in both the figures correspond to the calculations from the 
non-interacting (ideal) HRG model in which the mass of the hadrons are same as their 
rest mass ($m_0$). In Fig~\ref{fig:fig2}a, the effective mass remains at their vacuum 
values up to $T\approx$ 150 MeV at lower baryon chemical potential (up to 
$\mu\sim$ 200 MeV). In case of higher $\mu \sim$ 500 MeV, the effective mass 
starts reducing 
even at lower $T\sim$ 70--80 MeV. In Fig~\ref{fig:fig2}b, the effective mass remains 
unchanged as the vacuum values up to $\mu\sim 400$ MeV for $T = $ 100 MeV and there 
is sudden decrease of effective mass with increase in $\mu$. For higher temperature 
($T\ge$ 150 MeV) there is 2--3\% reduction of effective mass upto $\mu =$ 200 
MeV, thereafter gradual decrease as a function of $\mu$. The choices of $T$ up 
to 150 MeV and $\mu=$ 500 MeV are taken, because these higher values of $T$ and 
$\mu$ can be reached in heavy-ion collision experiments. In the lower two
panels Fig.~\ref{fig:fig2}c and Fig.~\ref{fig:fig2}d shows the reduced mass 
ratios for 
baryons and hyperons as a function of $T$ and $\mu$, respectively. The effective mass 
of baryons reduces significantly with increasing temperature and baryon density. This 
reduction is due to increase in attractive force between baryons in the nuclear 
matter as a function of temperature and baryon density. However, in the ideal HRG 
model, the masses of baryons remain constant due to the absence of interaction 
between the hadrons. Recently in the hadron resonance gas model~\cite{Sasaki:2019cqk}, it has been investigated the fluctuations and correlations involving 
baryon number in hot hadronic matter considering the in-medium mass 
effect of  baryons. The effective mass is used in the rest of the present study.

As discussed in the previous section about the value of hyperon couplings 
for scalar and vector mesons, we have constructed a volume independent quantity, 
the ratios of the yields of protons to that of $\Lambda + \Sigma^0$ baryons. 
Figure~\ref{fig:fig2_1} shows the ratios as a function of $T$ for both ideal and 
interacting HRG models. The ratio provides an useful diagnostic for the particle
content, hence the interaction strengths, in the baryon sector. The interacting
model shows an increase in the ratios relative to the ideal HRG case. Within the
HRG model, at around $T$ = 150--160 MeV, the ratio is slightly above 2.0, and 
the interacting HRG model leads to further increase in the ratio. However, the
experimental data obtained by ALICE collaboration on (multi)strange particle
production in Pb$+$Pb collisions at 
$\sqrt{s_{_{NN}}} =$ 2.76 TeV ~\cite{ALICE:2013mez,ALICE:2013cdo,ALICE:2013xmt} 
suggests that the value is around 1.3, which is significantly lower than the 
HRG result, and also goes against the trend demonstrated by the iHRG model. 
The value of 1.3 in the experimental data may be explained by: a suppressed 
proton yield~\cite{Andronic:2018qqt} with non-resonant interaction and an 
increase in hyperon yield. Further, the non-resonant interaction can  include dynamical features 
like roots, and coupled-channel effects~\cite{Lo:2020phg}. These effects are absent in the present 
RMF base iHRG model, and could be a reason why results from iHRG model displayed the opposite
 trend to data.
 
Figure~\ref{fig:fig3} shows the thermodynamic parameters ($p/T^4$, $\varepsilon/T^4$, 
$s/T^3$, $C_V/T^3$, $c_s^2$, and $(\varepsilon-3p)/T^4$) as a function of temperature 
for different $\mu$ = 0, 200 MeV, and 500 MeV. These parameters are calculated 
with ideal case of the HRG model. It is observed that the values of 
thermodynamic parameters as a function of $T$ are the same for any value of $\mu$ up 
to 200 MeV. However, there is a significant difference at higher baryon chemical 
potential. This indicates, when the baryon density increases along with temperature, 
the system is more excited and the production of particles are more as a result, the 
pressure and energy density increase as a function of temperature. However, the 
square of speed of sound decreases as a function of temperature, which implies that 
the rate of change of energy density is faster than the pressure due to more excited 
thermal energy and more number of degrees of freedom. At higher temperature, the 
square of speed of sound for all three chemical potential tends to similar values. 
That may be the region where the nuclear matter changes its phase structure.

Figure~\ref{fig:fig4} shows the thermodynamical parameters ($p/T^4$, $\varepsilon/T^4$, 
$s/T^3$, $C_V/T^3$, $C_s^2$, and $(\varepsilon-3p)/T^4$) as a function of chemical 
potential for different $T$ = 100 MeV, 150 MeV and 200 MeV. These parameters are 
calculated with ideal case of HRG mode same as shown in Fig.~\ref{fig:fig3}. It is 
observed that for different values $T$ (100 MeV, 150 MeV and 200 MeV), there is a 
systematically increase in the values of thermodynamic parameters as a function
of $\mu$ for different temperatures. However, in case of the 
square of speed of sound for $T$ = 100 MeV and 150 MeV, the change is very 
insignificant as a function of chemical potential or high baryon density. It clearly 
says that at around $T\sim$ 150 MeV and above, the system is in another state of 
matter, may be a phase transition to exotic matter, the proportionate change of 
pressure with respect to energy density is the same as a function of chemical 
potential.

\begin{figure*}[ht]
\bc
\includegraphics[width=1.0\textwidth]{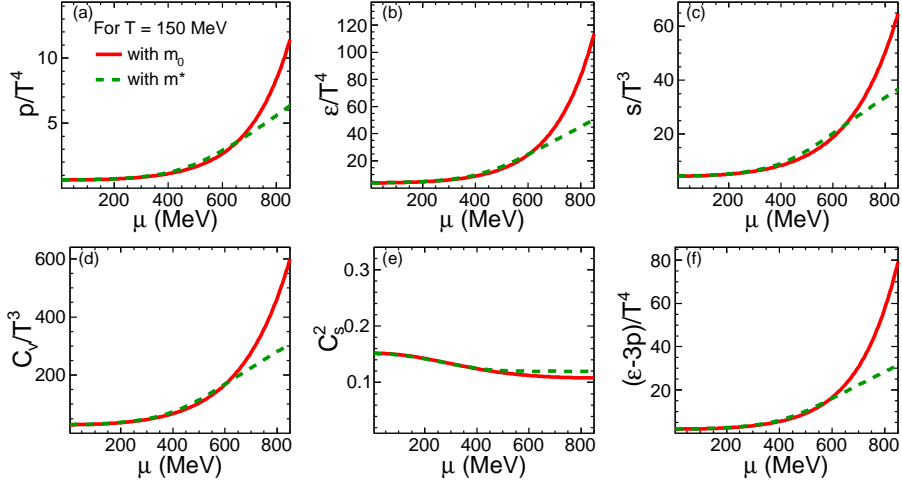}
\caption{The equation of states of interacting HRG (iHRG) as a function of 
$\mu$ at $T=$ 150 MeV. The solid red line is for HRG calculations using $m = 
m_0$ and the dashed green line is for iHRG calculations using $m = m^*$.} 
\label{fig:fig7}
\ec
\end{figure*}

\begin{figure*}[!ht]
\bc
\includegraphics[width=1.0\textwidth]{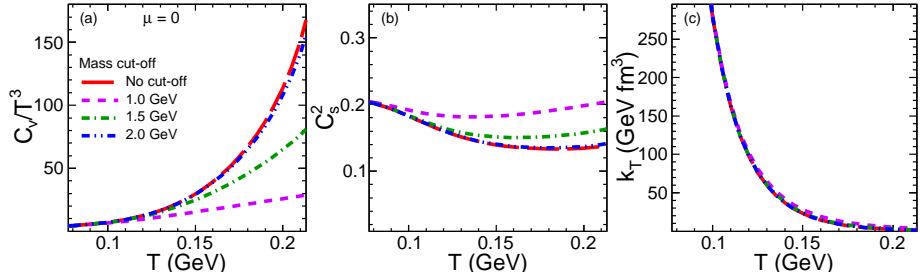}
\caption{The specific heat $C_V/T^3$ (panel (a)), speed of sound $c_s^2$ 
(panel (b)), and the isothermal compressibility $k_T$ (panel (c)) as a function 
of $T$ for inclusion of particles of different mass cut-offs at $\mu=$ 0 using 
iHRG model.} 
\label{fig:fig9}
\ec
\end{figure*}
\begin{figure*}[ht]
\bc
\includegraphics[width=1.0\textwidth]{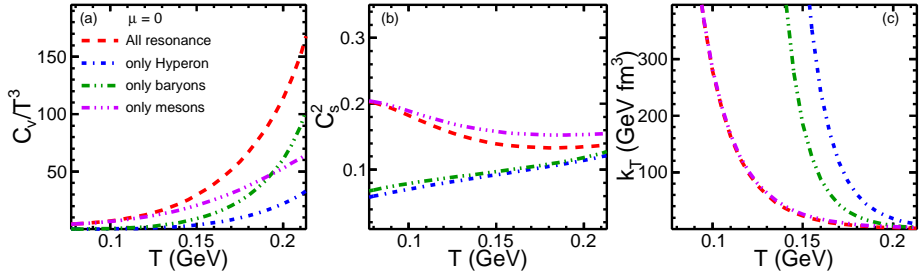}
\caption{Variation of $C_V/T^3$ (panel (a)), $c_s^2$ (panel (b)), and the $k_T$ 
(panel (c)) as a function of $T$ by taking different type of particles: all 
resonances (dashed red line), only hayperons (dashed-dotted blue line), only 
baryons (dashed-double dotted green line), and only mesons (dashed-triple 
dotted magenta line) at $\mu=$ 0 using iHRG model.} 
\label{fig:fig10}
\ec
\end{figure*}
\begin{figure*}[ht]
\bc
\includegraphics[width=1.0\textwidth]{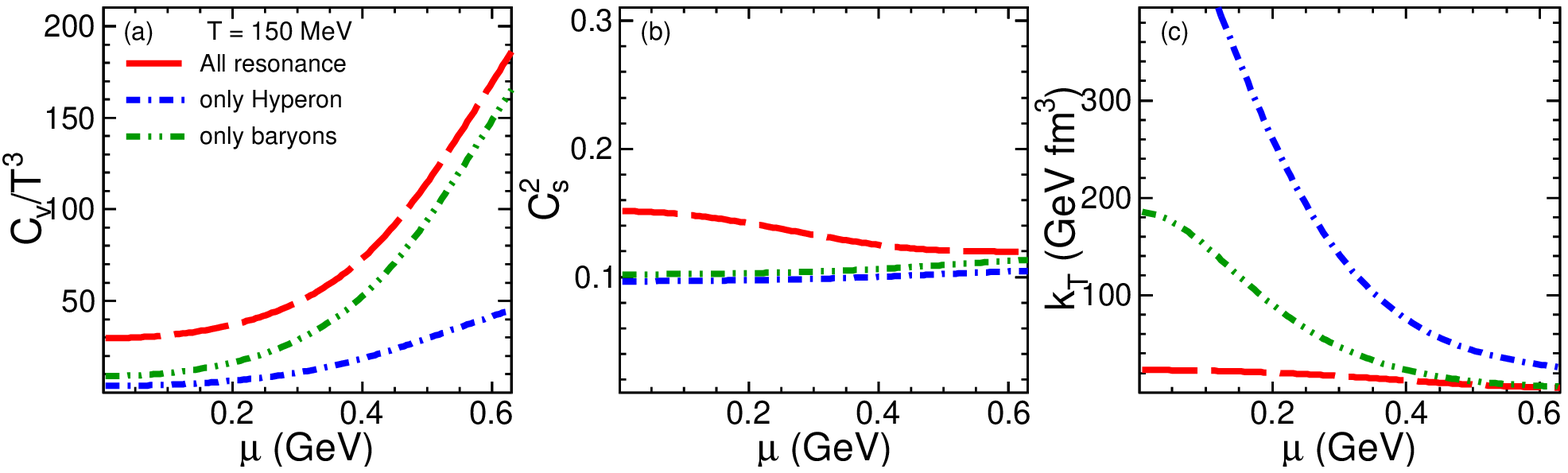}
\caption{Variation of $C_V/T^3$ (panel (a)), $c_s^2$ (panel (b)), and the $k_T$ 
(panel (c)) as a function of $\mu$ by taking different type of particles: all 
resonances (dashed red line), only hayperons (dashed-dotted blue line), and 
only baryons (dashed-double dotted green line) at $T=$ 150 MeV using iHRG 
model.} 
\label{fig:fig11}
\ec
\end{figure*}

The thermodynamical parameters ($p/T^4$, $\varepsilon/T^4$, $s/T^3$, $C_V/T^3$, 
$c_s^2$, and $(\varepsilon-3p)/T^4$) as a function of temperature at zero chemical 
potential calculated using HRG and iHRG models are shown in Fig.~\ref{fig:fig5}. The 
HRG and iHRG results are compared with the lattice QCD results of the Hot-QCD 
Collaboration~\cite{Bazavov:2014pvz} and the Wuppertal-Budapest (WB) 
Collaboration~\cite{Borsanyi:2013bia} shown in the same figure. The solid red line is 
for the HRG model and the blue dashed line is for iHRG model. The lattice data are 
very well described by both the models up to $T\sim$ 180  MeV. In case of $C_V/T^3$ 
and $c_s^2$ data in both models explain up to $T \le$ 170 MeV temperature range. It 
is very interesting to note that the iHRG model starts showing the difference from 
the HRG model calculations at temperature above 170 MeV in all thermodynamical 
parameters. However, in the case of $c_s^2$, both the models show almost no 
difference as a function of temperature. The equation of state, specifically the 
pressure at a higher temperature and at around the QGP region is more, as a result, 
the flow of particles from the collisions is more. It is noticed that to explain 
lattice data, the pressure has to be more harder or more repulsive force due to 
vector mesons is required at around temperature above 170 MeV.

Fig.~\ref{fig:fig6}, shows the same thermodynamical parameters as a function of 
temperature for different hadron species, where there are all particles using ideal 
HRG (solid red line), all hadrons (dashed-dotted cyan line), only mesons (dashed 
orange line), mesons plus nucleons (dashed blue line) and mesons plus nucleons and 
deltas (dashed-dotted green line) at zero chemical potential in iHRG models. The 
results from these models are compared with lattice calculations.  The lattice data 
can be well described by both models up to $T\sim$ 160 MeV. However, for models with 
mesons plus nucleons and mesons plus nucleons and deltas are well-matched with data 
up to temperature 200 MeV within the error bar, except for the square of the sound 
speed.This indicates that, the temperature above 160 MeV, which is close to the 
transition temperature to QGP phase, the production of particles are mainly mesons, 
nucleons and deltas due to Boltzmann suppression in masses.
Since the net baryon number (baryons minus anti-baryons) at 
around high temperature ($T\ge $ 150 MeV) and very high center of mass of 
collision 
energies, is less, therefore, other heavy baryons may be less around those regions. 
Another interesting point is noticed in lattice data that at around and above 
temperature 160 MeV, the speed of sound square shows an increasing trend. However, a 
slowly decreasing trend is seen in the cases of meson with nucleon and mesons with 
nucleon and delta baryons, that indicates that the interaction between the particles 
is soft as compare to lattice results. 

In Fig.~\ref{fig:fig7}, we compare  the thermodynamical parameters as a function of 
chemical potential at fixed temperature of 150 MeV in both HRG and iHRG models. The 
thermodynamic parameters from both the models are in good agreement for $\mu$ up to 
650 MeV, thereafter, there is a significant difference from iHRG to HRG due to 
strong attractive force, that reflects in the reduction in the mass of the baryons.
\begin{figure*}[!ht]
\bc
\includegraphics[width=1.0\textwidth]{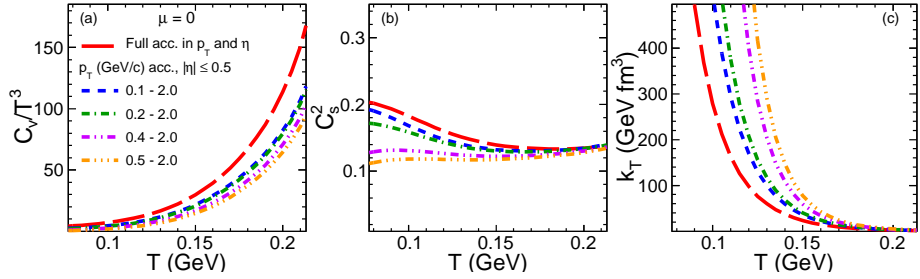}
\caption{The $C_V/T^3$ (panel (a)), $c_s^2$ (panel (b)), and $k_T$ (panel (c)) 
as a function of $T$ for different $p_T$ acceptances within $|\eta|\le$ 0.5 at
$\mu=$ 0 using iHRG model. Calculation for full $p_T$ and $\eta$ acceptance 
are shown (long dashed red line) for comparison.} 
\label{fig:fig12}
\ec
\end{figure*}
\begin{figure*}[ht]
\bc
\includegraphics[width=1.0\textwidth]{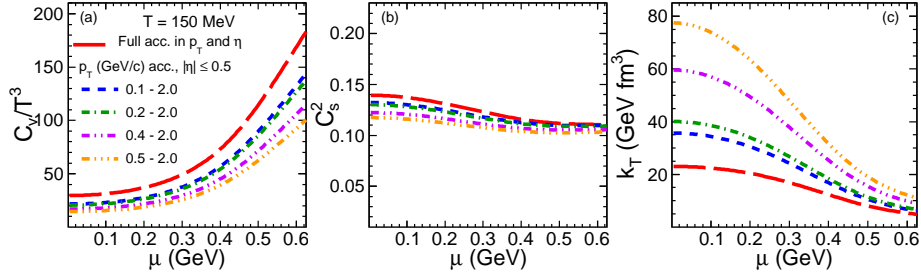}
\caption{Thermodynamic quantities same as in Fig.~\ref{fig:fig12} as a function 
of baryon chemical potential $\mu$ for different $p_T$ acceptances within 
$|\eta|\le$ 0.5 at $T=$ 150 MeV using iHRG model. Calculation for full $p_T$ and 
$\eta$ acceptance are shown (long dashed red line) for comparison.} 
\label{fig:fig13}
\ec
\end{figure*}

Figure.~\ref{fig:fig9} shows the $C_V/T^3$, $c_s^2$, and $k_T$ values are shown as a 
function of temperature for different higher mass cut-off at $\mu = $ 0 in the 
iHRG model. When higher mass particles are included ($m_0 \ge$ 1.0 GeV 
cut-offs), that means, increasing the degrees of freedom in the system, thus the 
$\varepsilon$ increases, as a result the $C_V$ (Fig.~\ref{fig:fig9}a) increases 
as a function of temperature, but in case of $c_s^2$ (Fig.~\ref{fig:fig9}b), 
with increase of mass cut-off then it decreases. At lower $T$, $C_V$ is 
dominated by contributions from mesons and at higher $T$ contributions to $C_V$ 
dominated by heavy baryons. It is clear from this observation that, the change 
of pressure due to the increase in temperature and increase of degrees of 
freedom, does not quantify significantly. This can also be 
verified from $k_T$ plot in (Fig.~\ref{fig:fig9}c) as a function of temperature, 
where $k_T$ does not change to each other with increasing mass cut-off values in the 
interacting HRG model. However, in the case of the ideal HRG case, it is observed 
that $k_T$ has a lower value when higher mass resonances added into the 
system~\cite{Khuntia:2018non}. The $k_T$ values decrease with temperature indicates 
that the matter formed is denser, hence it is difficult to compress the system. The 
temperature around phase transition from hadronic to partonic phase, the values of 
$k_T$ vanish with density, due to the super fluid nature of the quark gluon matter.
\begin{figure*}[ht]
\bc
\includegraphics[width=1.0\textwidth]{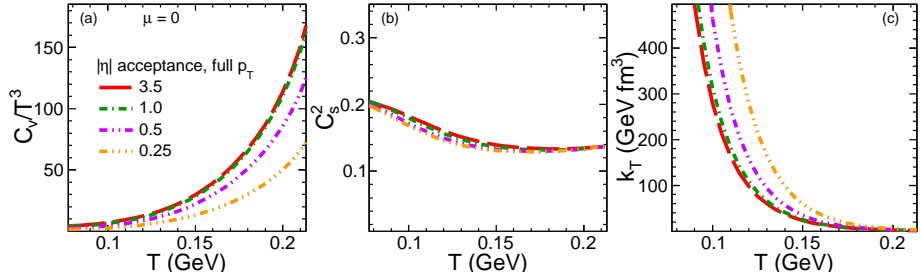}
\caption{The $C_V/T^3$ (panel (a)), $c_s^2$ (panel (b)), and $k_T$ (panel (c)) 
as a function of $T$ for different $\eta$ acceptances with full $p_T$ 
acceptance at $\mu=$ 0 using iHRG model.} 
\label{fig:fig14}
\ec
\end{figure*}
\begin{figure*}[t]
\bc
\includegraphics[width=1.0\textwidth]{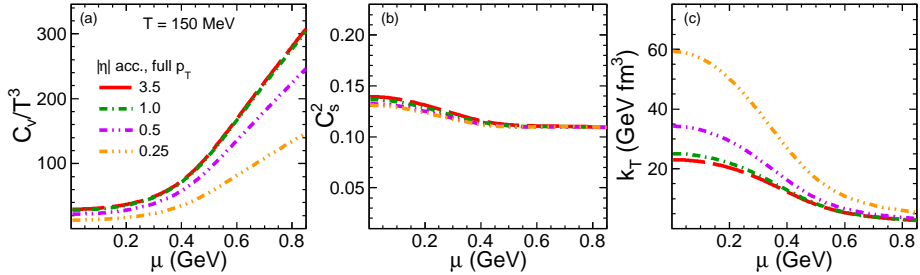}
\caption{Thermodynamic quantities same as in Fig.~\ref{fig:fig14} as a function 
of baryon chemical potential $\mu$ for different $\eta$ acceptances with full 
$p_T$ acceptance $T=$ 150 MeV using iHRG model.} 
\label{fig:fig15}
\ec
\end{figure*}

In Fig.~\ref{fig:fig10}, $C_V/T^3$, $c_s^2$, and $k_T$ values are plotted as a 
function of temperature at zero chemical potential for different species of the 
particles in the iHRG model. With the increasing number of degrees of freedom, the 
total energy density increases as a function of temperature, which can be very much 
evident in the Fig.~\ref{fig:fig10}a. At $T\sim$ 190 MeV, the green dashed line 
with only baryons takes over the magenta dashed line for only mesons, it means 
that the baryons energy density is much more than the mesons due to their heavy 
mass at such high temperatures. In the case of  Fig.~\ref{fig:fig10}b, the 
baryon and hyperon lines are much less than all resonances (red dashed line) and 
all mesons (magenta line). Since the baryons are heavier than mesons and the 
effect of baryons are much less in medium compare to mesons, therefore, the 
square of velocity of sound is much 
less at a lower temperature, however, all resonances, mesons and strange and 
non-strange baryons tend to vary similar square of velocity of sound at high 
temperature  ($\ge$ 200 MeV). If one observes $k_T$ value with different species, it 
is clear that the $k_T$ is very small for all the cases at $T \ge$ 200 MeV. The 
baryons start appearing first and then hyperons in the hadronic matter, therefore, 
the value of $k_T$ for baryons appear first than hyperons at high temperature, as is 
shown in Fig.~\ref{fig:fig10}c. However, the $k_T$ value starts appearing at  early 
value of temperature and reaches close to zero for baryons than the hyperons. After 
the temperature more than 200 MeV, the $k_T$ vanishes, which means the matter is in 
the QGP state, which is in a perfect fluid state. A similar trend is also seen 
in 
Fig.~\ref{fig:fig11}. Here $C_V/T^3$, $c_s^2$, and $k_T$ values are plotted as a 
function of chemical potential at fixed temperature 150 MeV in iHRG model. Since 
there is very less number of hyperons at $T=$ 150 MeV and at low chemical potential 
or low baryon density, the value of $C_V/T^3$ is less and increases with chemical 
potential. The $C_V/T^3$ value is high for all resonances at low chemical due to the 
large number of degrees of freedom for all particles. The intermediate value for 
$C_V/T^3$ is due to only baryons as seen in the figure includes hyperons and 
non-strange baryons. Similarly, the speed of sound for all resonance is higher than 
the only baryons and only hyperons as seen in Fig.~(\ref{fig:fig11}b). The $c_s^2$ 
value is very less dependent on $\mu$ at $T=$ 150 MeV for only hyperons and only for 
baryons. In the case of all resonances, the velocity of sound square is high at low 
density and merges to values of only baryons at high density with fixed quark-gluon 
plasma temperature $T=$ 150 MeV. In Fig.~\ref{fig:fig11}c, the $k_T$ value is much 
less for all resonances as compared to considering only baryons or only hyperons as a 
function of chemical potential for fixed temperature 150 MeV. We notice here that the 
mass of baryons and hyperons is higher than resonance particles, therefore, the 
$k_T$ value is high and slowly goes to zero at high chemical potential and hence 
high baryon density. So it is hard to compress thermally the matter at low density 
for hyperons and baryons than high density at a fixed temperature around the 
quark-gluon transition temperature region.

There are limitations in the experimental measurements, which are a fraction of the 
total available phase-space. Also, a different experiment may have different 
kinematic acceptance~\cite{Garg:2013ata}. Hence, it is important to study the 
$C_V/T^3$, $c_s^2$, and $k_T$ as a function of $p_T$ and $\eta$ acceptance. 
Figures~\ref{fig:fig12}--\ref{fig:fig15} show the $C_V/T^3$, $c_s^2$, and $k_T$ as a 
function of chemical potential at fixed $T$ and as a function of temperature at 
$\mu=$ 0 for different $p_T$ and $\eta$ acceptance range so that the model 
calculations can be compared with experimental heavy-ion collision data at RHIC 
and LHC energies. Figures.~\ref{fig:fig12},~\ref{fig:fig13} show the $T$ and 
$\mu$ dependence above three thermodynamic quantities, respectively using iHRG 
model. The values for full acceptance in $p_T$ and $\eta$ are shown in the red 
dashed lines and other $p_T$ acceptance cuts are also shown with $|\eta| <$ 0.5. 
It is observed that there is clear $p_T$ acceptance dependence for $C_V/T^3$ and 
$k_T$ as a function of both $\mu$ and $T$. While applying higher momentum cuts, 
the $C_V$ and $k_T$ show the opposite trend as compared to lower momentum cuts. 
In the case of $c_s^2$, the acceptance effect is observed at lower $\mu$ and $T$ 
values. Also, we noticed that with increasing chemical potential or baryon 
density, the $k_T$ value decreases and merges each other for fixed temperature, 
where the QGP transition is expected.

\begin{figure}[!ht]
\bc
\includegraphics[width=0.8\textwidth]{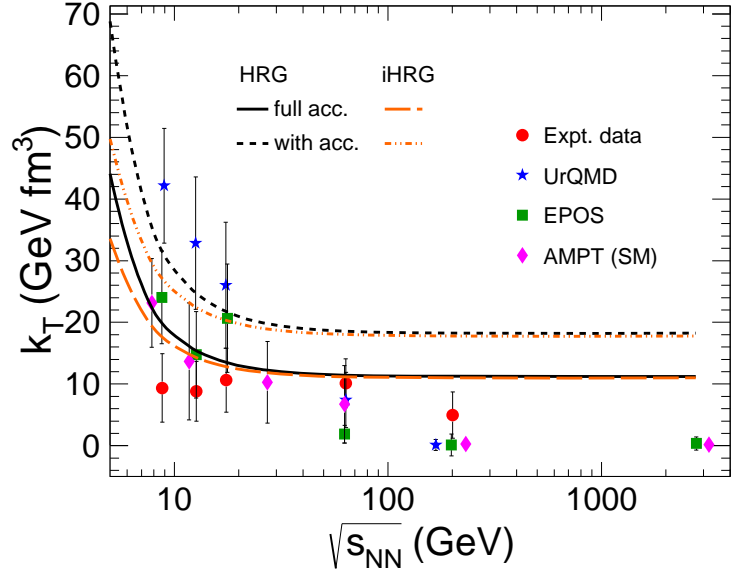}
\caption{The collision energy dependence of isothermal compressibility $k_T$ 
calculated using ideal HRG and iHRG models. The model calculations for both 
full acceptance and limited acceptances are also shown. The transport model 
calculations from UrQMD, AMPT, and EPOS along with other experimental 
measurements are shown for comparison (data are taken 
from~\cite{Mukherjee:2017elm} and references therein.).} 
\label{fig:fig16}
\ec
\end{figure}
Similar plots of the $C_V/T^3$, $c_s^2$, and $k_T$ as a function of chemical 
potential at fixed  $T=$ 150 MeV and as a function of temperature for $\mu=$ 0  with 
full $p_T$ acceptance range and different $\eta$ acceptance ranges are shown in 
Fig.~\ref{fig:fig14} and  Fig.~\ref{fig:fig15}. The nature of curve in 
Fig.\ref{fig:fig14}(a, c)  and  Fig.~\ref{fig:fig15}(a, c) are very similar to in 
Fig.\ref{fig:fig12}(a, c)  and  Fig.~\ref{fig:fig13}(a, c). However, the $c_s^2$ 
values in both the figure Fig.\ref{fig:fig14}b and  Fig.~\ref{fig:fig15}b do not 
change much and there is no $\eta$ dependence observed for $c_s^2$. This we may 
interpret as, at temperature close to QGP transition, the matter is having less 
interaction at lower density. 

In order to compare with the experimental measurements, the parameterization of 
freeze-out parameters $\mu$ and $T$ as a function of the center of mass-energy 
(\sqsn) are used as given in Ref.~\cite{Cleymans:2005xv}. Figure~\ref{fig:fig16} 
shows the isothermal compressibility as a function of center of mass energies 
calculated using both for HRG and iHRG models along with results from other heavy-ion 
transport models and experimental data. The experimental data, red dots are from 
lower energy to high energy. The other models, UrQMD, EPOS, and AMPT (SM) are denoted 
as blue star, green square, and magenta diamond symbols, respectively. These model 
results are taken from~\cite{Mukherjee:2017elm} and references therein. The HRG and 
iHRG model calculations are shown for full $p_T$ and $\eta$ acceptance and $0.2\le 
p_T\le 2.0 $ GeV/$c$ and $|\eta| \le 0.5$ acceptances. The effect of acceptance is 
clearly observed for $k_T$ as a function of \sqsn. We have noticed that calculations 
of both HRG and iHRG models for full $p_T$ and $\eta$ acceptance are close to the 
data from experiments as well as from other models. However, with selected $p_T$ and 
$\eta$ cut-off, the calculated values in both the models are  close to other models 
up to center of mass-energies 20 GeV but overestimate the experimental data for 
all energies. At low energies up to 10 GeV, the calculations from iHRG model 
is close to data than HRG model, that is because of the inclusion of interaction 
between baryons, which is significant at low energies. At high energies, the iHRG and 
HRG model are very close to each other, there is hardly any difference after 40 GeV 
of center of mass-energies. From these calculations, it is observed that, isothermal 
compressibility decreases rapidly with increasing \sqsn up to SPS energies ($\sim 10$ 
to 20 GeV) and remains constant thereafter up to LHC energy. This change of behavior 
of $k_T$ from lower energy to higher energy could be the onset of phase transition 
from hadron gas to partonic matter.
\section{Summary}
\label{sec:summary}
In summary, we have studied the thermodynamical quantities at finite temperature with 
vanishing $\mu$ and at finite chemical potential with $T=$ 0 using ideal HRG and 
interacting HRG (iHRG) models. The interaction is introduced through RMF in the HRG 
model. The interaction terms are based on the MFT among the baryons through meson 
exchange potential. The attractive interaction is realized by introducing the scalar 
meson exchange and the repulsive interaction among the baryons are introduced through 
vector meson exchange. The iHRG model is able to reproduce the lattice QCD equation 
of state at $\mu=$ 0 with finite temperature and at zero temperature with finite 
chemical potential or finite baryon density. At finite baryon density with zero 
temperature, the properties of nuclear matter at saturation density is well explained with a particular set of parameters as well.

In the MFT model, we have taken stable parameters to merge with the HRG model, 
called as STO-5, which describes well to infinite nuclear matter and data of 
finite nuclei. There is a substantial effect of interaction on the 
thermodynamical quantities at higher $\mu$ and $T$ values in the iHRG model. 
Also, the hyperon baryons are taken with proper interaction based on the MFT 
model. We have taken different options in the iHRG model taking all particles, 
only baryons, only hyperon baryons, and only resonances.

We have constructed a volume independent quantity, the ratios of the 
yields of protons to that of $\Lambda + \Sigma^0$ baryons which provides 
an useful diagnostic for the particle content, hence the interaction strengths, 
in the baryon sector. Within the HRG model around $T$ = 150--160 MeV, the 
ratio is slightly above 2.0, and the interacting HRG model leads to further 
increase in the ratio. However, the experimental result, which is around 1.3 
significantly lower than the HRG result, and also goes against the trend 
demonstrated by the iHRG model. Further, the non-resonant interaction can 
include dynamical features like roots, and coupled-channel effects. These 
effects are absent in the present RMF base iHRG model, and could be a reason 
why results from iHRG model displayed the opposite trend to data. This could 
be a good motivation for future work on re-examining many of the model 
assumptions in the RMF model, in comparison to the approach using empirical 
phase shift.

We have calculated the specific heat, thermal compressibility as well as square of 
speed of sound using iHRG model as a function of $\mu$ and $T$. The effect of mass 
cut-off, effects of including hyperons, baryons, and mesons are also presented which 
show a large effect on the studied quantities. The effect of kinematic acceptances 
($p_T$ and $\eta$) are studied to compare with the experimental measurements both as 
a function of $\mu$ and $T$. The $C_V$ and $k_T$ values are affected by the kinematic 
consider the acceptance effect while comparing the theoretical models with 
experimental measurements. The isothermal conductivity is studied as a function of 
collision energies, which decreases rapidly with increase in \sqsn up to SPS energy 
and remains constant thereafter up to LHC energy. This study provides reference 
baseline for comparison with the experimental data using purely thermal model and 
with the inclusion of interaction.

\end{document}